\documentclass[pra,twocolumn,amsmath,amssymb,superscriptaddress]{revtex4-2}
\usepackage{graphicx,hyperref,enumerate}
\usepackage[T1]{fontenc}%without this, Polish diacritics in the bibliography section throws errors
\usepackage{mathtools}

\allowdisplaybreaks

\usepackage{amsthm}
\usepackage{float}

\newcommand{\arXiv}[1]{\href{http://www.arXiv.org/abs/#1}{arXiv:#1}}
\renewcommand{\doi}[2]{\href{http://dx.doi.org/#2}{#1}}
\newcommand{\beq}{\begin{equation}}
\newcommand{\eeq}{\end{equation}}
\newcommand{\del}{\partial}
\newcommand{\dsty}{\displaystyle}
\newcommand{\ssty}{\scriptstyle}
\newcommand{\nn}{\nonumber}
\newcommand{\lambd}{\varepsilon}
\newcommand{\al}{\alpha}

\begin{document}

\title{Phase-space localization at the lowest Landau level\vspace{2mm}}

\author{Ben Craps} 
\affiliation{TENA, Vrije Universiteit Brussel (VUB) and International Solvay Institutes, Brussels, Belgium}
\author{Marine De Clerck}
\affiliation{DAMTP, University of Cambridge, Cambridge, United Kingdom}
\author{Oleg Evnin}
\affiliation{High Energy Physics Research Unit, Faculty of Science, Chulalongkorn University, Bangkok, Thailand}
\affiliation{TENA, Vrije Universiteit Brussel (VUB) and International Solvay Institutes, Brussels, Belgium}
\author{Maxim Pavlov}
\affiliation{TENA, Vrije Universiteit Brussel (VUB) and International Solvay Institutes, Brussels, Belgium}

\begin{abstract}
\rule{0mm}{8mm}We consider bosons with weak contact interactions in a harmonic trap
and focus on states at the lowest Landau level. Motivated by the known
nontrivial phase-space topography of the energy functional of the
corresponding Gross-Pitaevskii equation, we explore Husimi distributions
of quantum energy eigenstates in the classical phase space of the Schr\"odinger field.
With interactions turned off, the energy levels are highly degenerate
and the Husimi distributions do not manifest any particular localization
properties. With interactions turned on, the degeneracy is lifted, and
a selection of energy levels emerges whose Husimi distributions are
localized around low-dimensional surfaces in the phase space.\vspace{10mm}
\end{abstract}

\maketitle

\section{Introduction}

Dynamical systems tend to explore all configurations permitted by their conservation laws, in relation to the concept of ergodicity  \cite{Eckhardt:1995,Serbyn:2021,Vikram:2022}.
Localization of the evolution on a surprisingly small set of configurations always stands out as a notable physical signature. 

It is especially intriguing when a dynamical system commonly appearing in contemporary theoretical and experimental research
manifests such special properties.
Here, we turn our attention to nonrelativistic bosons with weak contact interactions confined by an external harmonic potential.
This setup is very common in studies of ultracold atomic gases \cite{Bloch:2008zzb,Fetter:2009zz}.
In a number of situations, including
rapid rotation, dynamics of such systems is dominated by the so-called Lowest Landau Level (LLL) sector \cite{Fetter:2009zz,Ho:2001zz,ABD,ABN}.
We shall focus on this sector, treated at linear order in the small coupling parameter, and observe a subset of quantum energy 
eigenstates whose corresponding phase-space distributions are well-localized.

In discussing phase-space localization, it is important to be clear which phase space we are talking about. Nonrelativistic many-body systems
admit equivalent first-quantized and second-quantized representations whose classical phase spaces are very different, resulting in either a mechanical system of particles or a nonrelativistic field theory, respectively. We shall focus on the second-quantized representation and its field-theoretic classical phase space of the corresponding (mean-field) Gross-Pitaevskii equation. It is in this setting that phase-space localization of some of the energy eigenstates in terms of their Husimi distributions \cite{Husimi,rev} will be observed. Studies of Husimi distributions are a common way to visualize the dynamical content of quantum eigenstates \cite{phasespaceloc1,phasespaceloc2,phasespaceloc3,phasespaceloc4,phasespaceloc5,phasespaceloc6,phasespaceloc7}. Our appeal to the phase space of the second-quantized representation is a departure from the common lore, however, and one of our key messages is that this strategy confers many advantages.

\section{Bosons with weak interactions and level splitting}

We start with the Hamiltonian of a self-interacting Schr\"odinger field $\Psi(x,y)$ in two dimensions in an external isotropic harmonic potential:
\begin{align}
 &\mathcal{H} =\mathcal{H}_0 + g \,\mathcal{H}_{int},  \label{Horg}\\
&\mathcal{H}_0= \frac{1}{2}\int \left[ \nabla \Psi^\dagger \cdot \nabla \Psi + (x^2 + y^2)\Psi^\dagger \Psi\right]\,dx\,dy,\\ 
&\mathcal{H}_{int}=\pi \int \,  \Psi^{\dagger2} \Psi^{2} \, dx\,dy.\label{Hint}
\end{align}
The canonical commutation relations 
\beq
[\Psi(t,x,y),\Psi^\dagger(t,x',y')]=\delta(x-x')\delta(y-y')\eeq 
are implied. We shall be interested in the regime of weak coupling, $|g|\ll 1$.

In the absence of interactions ($g=0$) the quantum system is immediately solvable.
After expanding $\Psi(x,y)$ through the 2d harmonic oscillator eigenfunctions $\psi_{nm}$ (where $n$ is the radial overtone and $m$ satisfying $|m|\le n$ is the angular momentum label) and interpreting the coefficients $a_{nm}$ as annihilation operators, one finds that the Fock states 
\beq
|\{\eta_{nm}\}\rangle=\prod_{nm}\frac{( a^\dagger_{nm} )^{\eta_{nm}}}{\sqrt{\eta_{nm}!}}|0\rangle
\label{Fockdef}
\eeq
form an orthonormal eigenbasis of $\mathcal{H}_0$ with energies
\beq
E_{\{\eta\}}=\sum_{nm} n \,\eta_{nm}.
\label{E0def}
\eeq
The integers $\eta_{nm}$ are the occupation number for modes $\psi_{nm}$.
The noninteracting spectrum is highly degenerate, since many Fock states \eqref{Fockdef} share the same energy \eqref{E0def}. 

A consequence of weak contact interactions is to lift these degeneracies by introducing energy shifts:
\beq
\tilde E_I=E+g\,\varepsilon_I.
\label{enshifts}
\eeq
The leading corrections $\varepsilon_I$ are found by application of the standard first-order degenerate perturbation theory:
one evaluates the matrix elements of (\ref{Hint}) between the unperturbed eigenstates \eqref{Fockdef} with the same energy \eqref{E0def} and diagonalizes the resulting matrix. The details of this computation can be found in \cite{split} and we reproduce them in Appendix~\ref{appLLL}. Energy levels of (\ref{Horg}-\ref{Hint}) and related systems have been frequently studied
\cite{split1,split2,split3,split3,split4,split5,split6,split7,split8,split9,split10}, though typically with a focus on small numbers of bosons. We are interested in the sector of the theory spanned by the Fock states \eqref{Fockdef} where the modes $a_{nm}$ with $n \neq m$ are turned off. This condition defines the LLL sector:
\beq
|\eta_0,\eta_1,\dots\rangle=\prod_{k=0}^\infty\frac{(a^\dagger_k)^{\eta_k}}{\sqrt{\eta_k!}}|0,0,0,\dots\rangle,\quad a_n \equiv a_{nn}.
\label{eq: LLL fock states}
\eeq
By contrast, in non-LLL states some of the modes with $n\ne m$ have nonzero occupation numbers. Because the LLL modes are located at the edge of the spectrum (there are no modes with $m>n$), it can be shown that $\mathcal{H}_{int}$ has vanishing matrix elements between LLL and non-LLL Fock states. For that reason, one can ignore the non-LLL modes completely for the purpose of computing the first-order energy shifts within the LLL sector. A detailed argument can be found in \cite{split}, and is also reproduced in Appendix~\ref{appLLL}. At the end of the day, the energy shifts $\varepsilon_I$ of LLL states are computed as the eigenvalues of the following LLL Hamiltonian:
\begin{align}
\mathcal{H}_{LLL}=&\hspace{1mm}{\textstyle\frac12}\hspace{-2mm}\sum_{n+m = k+l } \hspace{-2mm}C_{nmkl}  {a}^{\dagger}_{n}{a}^{\dagger}_{m} {a}_{k}{a}_{l},
\label{eq: LLL Hamiltonian}
\end{align}
with
\begin{align}
C_{nmkl} = \frac{(n+m)!}{2^{n+m}\sqrt{n!m!k!l!}}.
\label{CLLL}
\end{align}
This is the quantum version of the classical LLL Hamiltonian describing the weak coupling dynamics of Bose-Einstein condensates in 2d isotropic harmonic traps \cite{GHT,GT,LLLres,LLL2,LLL3}.
The LLL Hamiltonian (\ref{eq: LLL Hamiltonian}) commutes with the operators
\begin{align}
&{N} = \sum {a}^\dagger_k {a}_k, \,\,{M} = \sum k\,{a}^\dagger_k {a}_k,\,\,
{Z} = \sum\sqrt{k+1} \,a^\dagger_{k+1} a_k
\end{align}
and $Z^\dagger$. The nonvanishing commutators of these operators are
\begin{align}
& [{M}, {Z}] = {Z}, \quad [{M}, {Z}^\dagger]=-{Z}^\dagger, \quad
 [Z,Z^\dagger]=-N.\label{NEZcomm}
\end{align}
Here, $N$ is the total particle number, $M$ is the angular momentum, while $Z$ boosts the center-of-mass angular momentum. (The center-of-mass motion in harmonic potentials separates from other degrees of freedom,
and the center-of-mass performs simple harmonic oscillations. This is an example of a `breathing mode.' General implications of breathing modes for weakly nonlinear dynamics are treated in \cite{breathing}.) Quantum numbers correspond to mutually commuting operators, which can be taken as $\mathcal{H}_{LLL}$, $N$, $M$ and $ZZ^\dag$.

We arrived at the Hamiltonian (\ref{eq: LLL Hamiltonian}) starting from the problem of finding energy corrections (\ref{enshifts}) to the degenerate levels (\ref{E0def}). Such problems amount to diagonalizing finite-dimensional matrices made of the matrix elements of (\ref{Hint}) between
states within the same degenerate unperturbed level. Even though $\mathcal{H}_{LLL}$ acts on an infinite-dimensional Fock space, its matrix elements split correspondingly into finite-dimensional blocks acting on the Fock states within the same unperturbed degenerate level.
Indeed, $\mathcal{H}_{LLL}$ has vanishing matrix elements between Fock states with different values of $N$ and $M$. For any given values of $N$ and $M$, there is only a finite number of Fock vectors available. (These vectors are indexed by partitions of $M$ into at most $N$ parts.) One can then straightforwardly diagonalize $\mathcal{H}_{LLL}$ within any such $(N,M)$-block, and obtain the corresponding energy shifts for the degenerate level with $N$ particles, $M$ units of angular momentum, and unperturbed energy (\ref{E0def}), also equal to $M$. (The property of splitting into independent finite-dimensional block is shared by all Hamiltonians of the form (\ref{eq: LLL Hamiltonian}), for any values of the mode couplings $C_{nmkl}$. Systematic studies of such Hamiltonians were initiated in \cite{quantres}.)

\section{The ladder states}

The energy shifts (\ref{enshifts}), obtained by diagonalizing $\mathcal{H}_{LLL}$ restricted to a particular $(N,M)$-block, are in general complicated real numbers. There is a sprinkling of rational values within this spectrum, however, suggestive of some solvable structure. This structure has been manifested in \cite{ladder}, resulting in a family of explicit, exact eigenstates. It is for these special states that we find strong phase-space localization properties.

We refer the reader to \cite{ladder} for technical derivations, and simply quote here the key results that can be independently verified by a variety of analytic or numerical means. (Considerations of \cite{ladder} have been generalized to relativistic field systems in \cite{largec}.)
Within each $(N,M)$-block with $2\leq M \leq N$, the state 
\begin{equation}
|\psi^{N}_M\rangle =  \sum^{M}_{m=0} \frac{(-1)^m}{\dsty m! N^m} Z^m Z^{\dagger m} | N-M,M,0,0,0...\rangle
\label{eq: unnormalized ladder states}
\end{equation} 
is an eigenstate of the Hamiltonian \eqref{eq: LLL Hamiltonian} with eigenvalue
\beq
\varepsilon^{(N,M)}= \frac{N(N-1)}{2}-\frac{NM}{4}.
\label{epskern}
\eeq
Moreover, since the $Z$-operator commutes with the Hamiltonian and raises $M$ by 1, one can use it to transport the states \eqref{eq: unnormalized ladder states} from blocks with lower $M$ to obtain extra eigenstates in the current block, with energies forming an evenly spaced ladder-like pattern
\beq
\varepsilon^{(N,M)}_m= \frac{N(N-1)}{2}-\frac{N(M-m)}{4}.
\label{epsladder}
\eeq

These ladder states have been understood \cite{ladder} as the microscopic origin for previously analyzed time-periodic phenomena in the classical (mean-field) theory \cite{LLLres}. One can consider the classical counterpart of \eqref{eq: LLL Hamiltonian}:
\begin{align}
H=&\hspace{1mm}{\textstyle\frac12}\hspace{-2mm}\,\sum_{n+m = k+l } \hspace{-1.5mm}C_{nmkl}\, \bar {\alpha}_{n}\bar{\alpha}_{m} {\alpha}_{k}{\alpha}_{l},
\label{eq: classical LLL Hamiltonian}
\end{align}
with interaction coefficients \eqref{CLLL}. Here, $\alpha_n$ are complex variables and their conjugate momenta are $i\bar\alpha_n$. The corresponding equations of motion possess solutions of the form
\begin{align}
\alpha_n(t)= \left( \frac{a(t)}{p(t)} \,n+b(t) \right) \frac{p(t)^n}{\sqrt{n!}},
    \label{eq: ansatz}
\end{align}
where the function $|p(t)|$ turns out to be time-periodic.
In \cite{ladder}, it was shown how to form coherent-like combinations of the quantum states corresponding to \eqref{epsladder} that precisely reproduce \eqref{eq: ansatz} in the semiclassical limit. In this paper, we will be mainly interested in coherent-like combinations of the lowest ladder states \eqref{eq: unnormalized ladder states}, which correspond to a subclass of the solutions \eqref{eq: ansatz} with $Z=0$ \cite{LLLres,ladder}.

\section{The LLL energy functional}

A key ingredient in the derivation of the states \eqref{eq: unnormalized ladder states} is the following operator \cite{ladder}:
\begin{equation}
B \equiv \mathcal{H}_{LLL} - \frac{N(N-1)}{2} + \frac{1}{4}\left( NM-Z Z^\dagger \right).
\label{eq: definition B}
\end{equation}
By writing $B$ in terms of creation-annihilation operators, one notices that it annihilates the state $| N-M,M,0,0,0\dots \rangle$, and consequently also the states \eqref{eq: unnormalized ladder states} and all their $Z$-images (note that $B$ commutes with $Z$ and $Z^\dagger$). The condition of being annihilated by $B$ and  $Z^\dagger$ is an effective way to define the ladder states (\ref{eq: unnormalized ladder states}), and it also fixes their energies as (\ref{epskern}). Acting on (\ref{eq: unnormalized ladder states}) with $Z$ one obtains further ladder states with energies (\ref{epsladder}). We shall focus below on the lowest ladder states (\ref{eq: unnormalized ladder states}) in each block, since the remaining ones are obtained from those by boosting the center-of-mass motion.

In \cite{ladder}, it was conjectured on the basis of numerical and partial analytic evidence that $B$ is a nonnegative operator, so that its expectation value is positive on all states, except for the ladder states that it annihilates.
This would also imply the corresponding bound in the classical theory (\ref{eq: classical LLL Hamiltonian}),
\beq
H\ge\frac{N^2}{2}-\frac{NM-|Z|^2}4,
\label{classineq}
\eeq
which was subsequently proved in \cite{Schwinte}. It was moreover shown that the configurations saturating the bound \eqref{classineq} are precisely of the form \eqref{eq: ansatz}. Since the configurations in (\ref{eq: ansatz}) are parametrized by 3 complex numbers $a$, $b$ and $p$, the constant energy surface for the energy value saturating the bound (\ref{classineq}) has at most 6 real dimensions. This is in striking contrast with generic constant energy surfaces in the infinite-dimensional phase space of $\alpha_n$, which are themselves infinite-dimensional. Turning back to the corresponding quantum system, the ladder state energies are slightly below the classical bound (\ref{classineq}) and so they are very close to the bottom of the peculiar low-dimensional phase-space valley dictated by (\ref{classineq}).
(The difference between the quantum and classical bound $B \geq 0$ is due to ordering ambiguities. The ladder state energies slightly violate the classical bound due to the normal-ordered form of the quantum Hamiltonian, which effectively involves a subtraction of some amount of energy.) One thus expects that the ladder states will be very constrained in the way they explore the infinite-dimensional phase space and that they will be localized around low-dimensional surfaces. In the remainder of our treatment, we shall show quantitatively that this indeed happens.

\section{Phase-space localization of Husimi distributions}

Our strategy will rely on Husimi distributions \cite{Husimi,rev}, widely used to develop phase-space pictures of quantum eigenstates \cite{phasespaceloc1,phasespaceloc2,phasespaceloc3,phasespaceloc4,phasespaceloc5,phasespaceloc6,phasespaceloc7}.
For a given state $\psi$, the Husimi distribution is defined by
\beq 
Q_{\psi}(\alpha) \equiv |\langle \psi|  \alpha \rangle|^2
\label{Eq_Husimi}
\eeq
where $|\alpha\rangle$ is the many-body coherent state
\beq
|\alpha\rangle \equiv  e^{\sum_{i} \alpha_i a_i^\dagger-\alpha^*_i a_i}|0\rangle= e^{-|\alpha|^2/2} e^{\sum_{i} \alpha_i a_i^\dagger}|0\rangle
\eeq
with $|\alpha|^2 \equiv \sum_i |\alpha_i|^2$. For $|\psi\rangle$ in the $(N,M)$-block, the modes $a_{n>M}$ are in their vacuum states, so it suffices to deal with a $2(M+1)$-dimensional (real) phase space made of $\al_0,\ldots,\al_M$.

Husimi distributions are closely related to the more familiar Wigner phase-space distributions. One issue with the latter is that they are not necessarily positive and do not provide genuine probability distributions. An integral convolution of the Wigner function with a Gaussian filter results in the Husimi distribution, which is by construction always positive, see \cite{HusWig}.

When the coupling is turned off ($g =0$), the energy eigenstates in the LLL sector are linear combinations of the Fock states \eqref{eq: LLL fock states} whose Husimi distributions are
\beq 
Q_{\eta}(\alpha) = |\langle \alpha| \eta \rangle|^2 = e^{ -|\alpha|^2} \prod_{i} \frac{|\alpha_i|^{2\eta_i}}{\eta_i!} .
\label{eq: husimi fock state}
\eeq
These distributions peak at $|\alpha_i|^2 = \eta_i$ and are constant under independent phase shifts of $\alpha_i$. The Husimi distribution of a typical Fock state is localized around a high-dimensional torus in phase space. (These are just the Kolmogorov tori of the trivially integrable linear noninteracting system.) For a random linear combination of Fock states at fixed (large) $N$ and $M$, the distribution typically spreads even further over a high-dimensional region constrained only by the conservation of $N$ and $M$.

When $g>0$, the degeneracies are lifted, and the new nondegenerate eigenstates are determined by the LLL Hamiltonian \eqref{eq: LLL Hamiltonian}. The Husimi distribution of a Hamiltonian eigenstate is generically expected to decay exponentially fast away from the classical phase-space hypersurface 
defined by the constant values of the classical conserved quantities corresponding to the chosen eigenstate \cite{decay1,decay2}. For the LLL system, this hypersurface is generally codimension 4 as energy eigenstates are simultaneous eigenstates of the operators $\mathcal{H}_{LLL}$, $N$, $M$ and $ZZ^\dagger$. 
There is one further salient feature of the Husimi distributions related to its symmetries. Symmetries of a Hamiltonian do not immediately translate into symmetries of the Husimi distributions. Indeed, the Hamiltonian itself induces a time-translation symmetry, but advancing along classical trajectories does not keep the Husimi distribution constant. However, specifically for conservation laws of the form $\hat{h} = \sum_n h(n) a^\dagger_n a_n$, the associated symmetry transformations $\alpha_n\to e^{i\theta h(n)}\alpha_n$ leave the Husimi distribution invariant. We provide a proof of this statement in Appendix~\ref{appsymm}. Since $N$ and $M$ in (\ref{NEZcomm}) are precisely of this form, the Husimi distribution remains constant over the surface of the two-dimensional torus defined by
\beq\label{Thdef}
\Theta(\alpha^{\mbox{\tiny start}};\theta,\eta):\,\,\alpha_n=e^{i(\theta+n\eta)}\alpha^{\mbox{\tiny start}}_n,\,\,\,\,\, \theta,\eta\in[0,2\pi),
\eeq
for any given starting point $\alpha^{\mbox{\tiny start}}$.

Ladder states have energies just below the classical bound (\ref{classineq}), and correspondingly we do not expect their Husimi distributions to stray far away from the classical configurations \eqref{eq: ansatz} that saturate this bound, parameterized by 6 real numbers coming from the complex variables $a$, $b$, $p$. The accessible configurations associated to a quantum state $|\psi\rangle$ are further constrained by its quantum numbers, which are the eigenvalues of $\mathcal{H}_{LLL} ,N,M$ and $ZZ^\dagger$. For a generic ladder state, only three of these are independent, yielding three constraints. In the case of the lowest ladder state where $ZZ^\dagger = 0$, however, the corresponding classical constraint is $|Z|^2=0$, which is equivalent to $Z=0$. This results in a total of four constraints and leaves a two-dimensional surface of accessible configurations. This is the minimal number of dimensions one could in principle have, given the constancy of Husimi distributions over the tori (\ref{Thdef}).

To confirm this picture, we investigate numerically the phase-space localization properties of ladder states, and contrast their Husimi distributions with those of generic states \cite{codes}. Our approach consists in exploring the phase space close to the shell $M_{\psi}$ of codimension 4 defined by classical conserved quantities being equal to the corresponding quantum numbers of the state $\psi$, 
\begin{align}
M_\psi:\,\,\,&H(\al_n)=\langle\psi|\mathcal{H}_{LLL}|\psi\rangle,\,\,\, |Z|^2(\al_n)=\langle\psi|ZZ^\dagger|\psi\rangle,\nn\\
&N(\al_n)=\langle\psi|N|\psi\rangle,\,\,\,M(\al_n)=\langle\psi|M|\psi\rangle,\label{Mpsidef}
\end{align}
since the Husimi function should be localized there for all states. We shall start by identifying a configuration $\alpha_{\max}$ where the Husimi distribution takes a particularly large value. This value will have to be replicated over the entire two-dimensional surface $\Theta(\alpha_{\max})$. We will then see that the Husimi function decays exponentially away from $\Theta(\alpha_{\max})$ for ladder states, while for generic states we encounter a landscape of appreciable Husimi distribution values spread out over $M_\psi$. (Note that the classical energy $H$ defining $M_{\psi}$ differs slightly from the quantum energy of $\psi$ due to ordering ambiguities. The relative discrepancy vanishes in the semiclassical limit $N \rightarrow \infty$. In practice, we determine the classical values defining the shell $M_\psi$ by evaluating the conservation laws at $\alpha_{\max}$.)

We need to find a sample cloud of configurations where the Husimi distribution values are substantial, and compare the resulting geometries of Husimi distributions for ladder and non-ladder states. Due to the high dimensionality of the phase space, judicious choices should be made in designing this sampling. If the cloud is not sufficiently spread out, it will overlook the key features of the Husimi distribution in areas falling outside its coverage. (One needs to move
away from the maxima of Husimi distributions at $\alpha_{\max}$ to appreciate the extent of their spreading.) If the cloud is too dispersed, an unrealistically large number of sample points will be required in a high-dimensional space to ever hit appreciable values of the Husimi distribution. Thus, we must generate a cloud that restricts the search to a useful region of phase space, without being too restrictive. In practice, after constructing the phase-space point 
$\alpha_{\max}$ where the Husimi distribution peaks by using optimization algorithms, we use it as the starting value for a Metropolis-Hastings sampling run that 
generates a cloud of points. To make the procedure effective, as the target probability distribution for this run we choose a Gaussian decaying away
from the classical hypersurface (\ref{Mpsidef}), reminiscent of the formulas seen in \cite{decay1,decay2},
\beq
\begin{split} \label{GaussTarget}
 P(\alpha_n) =&\exp\Big[-\frac{\ssty\left( H(\alpha_n) - H_{\max} \right)^2}{\ssty 2\sigma_H^2}-\frac{\ssty\left(N(\alpha_n) - N_{\max}\right)^2}{\ssty 2\sigma_N^2} \\  
&-\frac{\ssty\left(M(\alpha_n) - M_{\max}\right)^2}{\ssty 2\sigma_M^2}-\frac{\ssty(|Z|^2(\alpha_n) - |Z_{\max}|^2)^2}{\ssty 2\sigma_{|Z|^2}^2}\Big].
\end{split}
\eeq
The subscript `$\max$' refers to the value of the corresponding functional on $\alpha_{\max}$.
For concreteness, we focus on a large block of dimension $5604$, with $N=70$ and $M=30$, and restrict our attention to eigenstates with the center-of-mass at rest, annihilated by $Z^\dagger$. The relevant ladder state (\ref{eq: unnormalized ladder states}) is the lowest energy state in the block. For our choice of a generic state, we pick energy level $5$ above the ground state. 
To compare the two states, we plot the Husimi function values of the sampled points against their distance $d$ to the two-dimensional surface $\Theta(\alpha_{\max})$. The distance of a phase-space point $\alpha$ to the surface $\Theta(\alpha_{\max})$ is defined as $d = \min_{x\in \Theta(\alpha_{\max})} |x-\alpha|^2$. Numerically, we find this distance by drawing a fine grid on $\Theta(\alpha_{\max})$, computing the distance of each point on the grid to $\alpha$. The distance $d$ is the minimum of this
\begin{figure}[H]
    \includegraphics[scale=0.5]{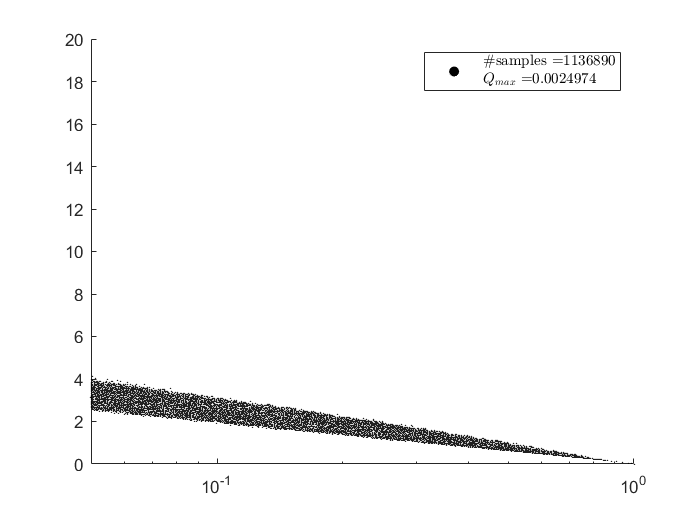}\\
    \includegraphics[scale=0.5]{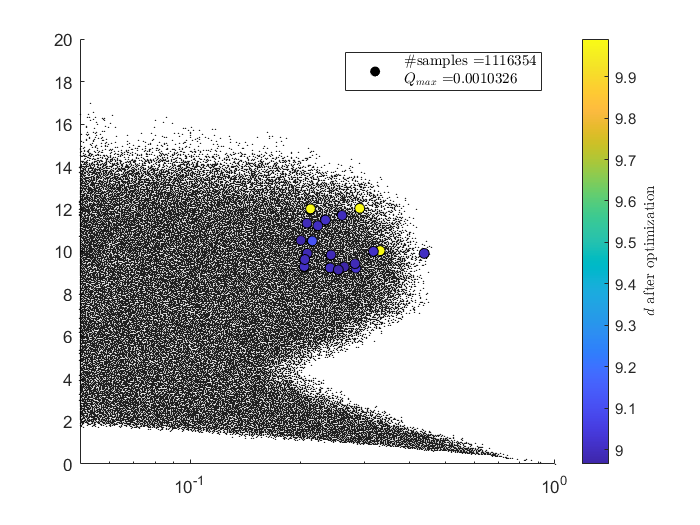}\\
   \begin{picture}(0,0)
     \put(160,220){$Q_\psi/Q_{\max}$}
     \put(160,20){$Q_\psi/Q_{\max}$}
     \put(10,150){$d$}
     \put(15,350){$d$}
   \end{picture}
    \caption{Husimi function values sampled over a cloud of phase-space points, plotted against their distance $d$ to the two-dimensional surface $\Theta(\alpha_{\max})$ where the largest Husimi function value lies. \textbf{Top:} for the ladder state with $|Z|=0$; \textbf{bottom:} for a non-ladder state with $|Z|=0$, both in the Hilbert space block with $N=70$ and $M=30$. The Gaussian distribution used in the sampling algorithm has mean values determined by the classical conservation laws at $\alpha_{\max}$, with $(N_{\max},M_{\max},Z_{\max},H_{\max})=(70,30,0,1925)$ (top) and $(N_{\max},M_{\max},Z_{\max},H_{\max})=(70,30,0,1947.04)$ (bottom), obtained by global optimization of the Husimi distribution starting from a randomly picked $\alpha_n$. The standard deviations of the Gaussian distribution used for sampling are set to $5\sqrt{2} \times (\sigma_N,\sigma_M,\sigma_{|Z|^2},\sigma_H) = (1,1,N,N)$ for both cases. For the non-ladder state, we further perform local optimization of the Husimi distribution for randomly selected points in the upper right area of the plot, and display the distance from the resulting local maximum to the initial two-dimensional surface, indicated by the color of the selected points. These results show that the Husimi distribution of the ladder state is localized on $\Theta(\alpha_{\max})$, while the region of high Husimi distribution spreads beyond $\Theta(\alpha_{\max})$ for generic states, with many distinct local maxima decorating the landscape.}
    \label{fig:ladder}
\end{figure}
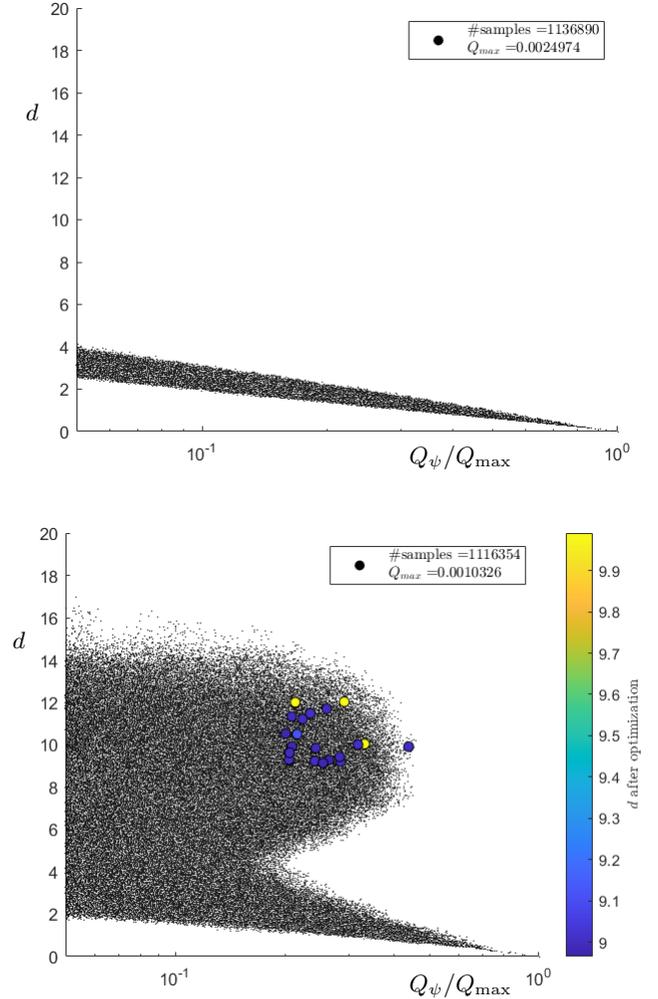
\noindent set.
The results are displayed in Fig.~\ref{fig:ladder} for the two chosen states. Some further numerical examples are given in Appendix~\ref{appsampl}, and they are consistent with the general picture presented here. Different choices for the standard deviations in \eqref{GaussTarget} have been tested to obtain an optimal sampling cloud. Clouds that are too dispersed would result in a large sampling of phase-space points with very small $Q_\psi$, while clouds that are too restrictive would not sample Husimi function values at distances far enough from $\Theta(\alpha_{\max})$, and would not be able to distinguish localized and non-localized states.

For the ladder state, Fig.~\ref{fig:ladder} shows an exponential correlation between the Husimi value and the distance away from $\Theta(\alpha_{\max})$. This reflects exponential decay away from the classically permitted region of motion, which coincides with $\Theta(\alpha_{\max})$ in this case. On the other hand, for
 the generic state, the bottom plot in Fig.~\ref{fig:ladder} shows a protrusion of points in the top-right direction where the Husimi function is large while being far from the two-dimensional surface $\Theta(\alpha_{\max})$. By performing a local optimization of $Q_\psi$ on phase-space points in this part of the figure, we discover new maxima of $Q_\psi$ that are located at significant distances away from $\Theta(\alpha_{\max})$.

\section{Conclusions}

We have confirmed that a selection of states with strongly localized phase-space distributions emerges when degenerate energy levels of free nonrelativistic bosons in a harmonic trap split due to weak contact interactions. The states with localized distributions either lie at the bottom of the fine structure emanating from the unperturbed degenerate levels, or are obtained by acting on the lowest states at a different free-boson level with the center-of-mass boost $Z$. While, classically, being close to the energy minimum guarantees being severely restricted in phase space, quantum-mechanically, the presence of a potential well does not guarantee even the existence of bound states, let alone their strong localization. Our analysis thus demonstrates that the energy valleys defined by the saturation of the bound (\ref{classineq}) are deep enough to provide for strong phase-space localization of the corresponding quantum energy eigenstates. 
It is an extra appealing feature that the structure of the LLL Hamiltonian can be exploited to give analytic expressions (\ref{eq: unnormalized ladder states}) for the wavefunctions of these special states \cite{extralddr}.
By contrast, if we move to levels with $M>N$, where the lowest energy eigenstates of the fine structure are not ladder states, the degree of localization decreases, which correlates intuitively with the energy gap (binding energy) at the bottom of the spectrum becoming more shallow. We provide the corresponding plots in Appendix~\ref{appnonladd}.

An unusual aspect of our treatment is that we have relied on the phase space of the classical Schr\"odinger field, corresponding to second-quantized quantum mechanics, rather than the particle phase space (positions and momenta) of individual bosons. This choice played an essential role in our treatment. Indeed, had we chosen to work in the first-quantized representation, we would have had to deal with the phase space of, say, 70 particles in 2 dimensions, with all directions playing a significant role, pushing the dimensionality beyond what one could realistically work with \cite{firstq}. By contrast, in the second quantized representation, the higher modes $\al_{n\gg 1}$ have very low occupation numbers and can be ignored in the analysis of phase-space distributions, leaving behind a problem of moderately high, manageable dimension.

We are not aware of prior treatments of the physics of ultracold atomic gases in the language of phase-space distributions of the second-quantized representation. 
This way of handling 	quantum states is very common, on the other hand, in quantum optics, also in conjunction with experiments. Indeed, the Schr\"odinger field plays the same role for identical nonrelativistic bosons that the Maxwell field plays for photons.
Quantum phase-space distributions in terms of the Maxwell field oscillator modes have been used ubiquitously in quantum optics, see \cite{Qopt} for a textbook exposition.
We have demonstrated the utility of this picture for visualizing the properties of energy eigenstates of weakly interacting bosons.
This approach should have further fruitful applications.

\section*{Acknowledgments:}

This work has been supported by Research Foundation Flanders (FWO) through project G012222N, and by Vrije Universiteit Brussel through the Strategic Research Program High-Energy Physics. The work of MDC is partially supported by STFC consolidated grants ST/T000694/1 and ST/X000664/1 and by the Simons Investigator award $\#$620869. OE is supported by Thailand NSRF via PMU-B (grant number B13F670063). MP is supported by the FWO doctoral fellowship 1178725N. The computational resources and services used in this work were provided by the VSC
(Flemish Supercomputer Center), funded by the Research Foundation Flanders (FWO) and
the Flemish Government. In the numerics, we have used
the partition generator \cite{part_gen}.

\appendix

\section{Derivation of the LLL Hamiltonian}\label{appLLL}

Here, we retrace the steps leading to the LLL system from the application of first-order perturbation theory to nonrelativistic bosons with weak quartic contact interactions in a 2d harmonic trap (following the derivation in \cite{split}):
\begin{align}
 &\mathcal{H} =\mathcal{H}_0 + g \,\mathcal{H}_{int},  \label{Horg_app}\\
&\mathcal{H}_0= \frac{1}{2}\int (  \nabla \Psi^\dagger \cdot \nabla \Psi + (x^2 + y^2)\Psi^\dagger \Psi)\,dx\,dy,\\ 
&\mathcal{H}_{int}=\pi \int \,  \Psi^{\dagger2} \Psi^{2} \, dx\,dy.\label{Hint_app}
\end{align}
We decompose the Schr\"odinger field $\Psi$ in terms of annihilation operators as 
\beq
\Psi(x,y)=\sum_{n,m} a_{nm} \psi_{nm}(x,y),
\label{psidecomp}
\eeq
where $\psi_{nm}$ solve the single-particle stationary Schr\"odinger equation
\begin{align}
\frac{1}{2}(-\del_x^2- \del_y^2 + x^2 + y^2)\psi_{nm}&=(n+1)\psi_{nm},\nonumber\\
-i\del_\varphi \psi_{nm}&=m\psi_{nm},
\end{align}
with $\varphi$ the angular coordinate in the $(x,y)$-plane. 

At $g = 0$, the spectrum is simple and highly degenerate with energies
\beq
E_{\{\eta\}}=\sum_{nm} n \,\eta_{nm}.
\label{E0def_app}
\eeq
The integers $\eta_{nm}$ are the occupation numbers of the Fock states made with $a_{nm}^\dagger$, which form an orthonormal eigenbasis for $\mathcal{H}_0$.  

When $ 0< g \ll 1$, the degeneracy splits due to the first order correction $\varepsilon_I$:
\beq
\tilde E_I=E+g\,\varepsilon_I.
\label{enshifts_app}
\eeq
The first step in finding the shifts $\varepsilon_I$ is to express $\mathcal{H}_{int}$ through creation-annihilation operators:
\begin{align}
\mathcal{H}_{int}=&\hspace{1mm}{\textstyle\frac12}\hspace{-0.5mm} \,\sum_{\mathclap{\substack{n_1,n_2,n_3,n_4 \geq 0\\m_1+m_2 = m_3+ m_4}}} \hspace{1mm}C_{n_1n_2n_3n_4}^{m_1m_2m_3m_4}  {a}^{\dagger}_{n_1m_1}{a}^{\dagger}_{n_2m_2} {a}_{n_3m_3}{a}_{n_4m_4},
\label{eq: interaction Hamiltonian}
\end{align}
where the constraint on the indices $m_i$ reflects conservation of angular momentum. 
The interaction coefficients in \eqref{eq: interaction Hamiltonian} are given by
\begin{equation}
C_{n_1n_2n_3n_4}^{m_1m_2m_3m_4} =2\pi \int \psi_{n_1m_1}^* \psi_{n_2m_2}^*
\psi_{n_3m_3}
\psi_{n_4m_4} r \, dr \, d\phi,
\label{intcoeff}
\end{equation}
where we adopt the same normalization for the eigenfunctions $\psi_{nm}$ as in \cite{split}.
Next, perturbation theory instructs us to consider the following matrix elements 
\beq
\langle \{\eta\}|\mathcal{H}_{int}| \{\eta'\}\rangle,
\label{blockdef}
\eeq 
where $|\{\eta\}\rangle$ and $|\{\eta'\}\rangle$ are Fock states with identical unperturbed energy $E_{\{\eta\}}$, and diagonalize the resulting matrix. Since $\mathcal{H}_{int}$ needs to be evaluated between two states with the same unperturbed energy, one finds that all terms in \eqref{eq: interaction Hamiltonian} with $n_1+n_2 \neq n_3+n_4$ are irrelevant for the purposes of first order perturbation theory and can be discarded. Hence, one can equivalently consider the Hamiltonian
\begin{align}
\mathcal{H}_{int}=&\hspace{1mm}{\textstyle\frac12}\hspace{-0.5mm}\,\sum_{\mathclap{\substack{n_1+n_2 = n_3+n_4 \\m_1+m_2 = m_3+ m_4}}} \hspace{1mm}C_{n_1n_2n_3n_4}^{m_1m_2m_3m_4}  {a}^{\dagger}_{n_1m_1}{a}^{\dagger}_{n_2m_2} {a}_{n_3m_3}{a}_{n_4m_4}.
\label{eq: updated interaction Hamiltonian}
\end{align}

In the main text, we are interested in the sector of the theory with maximal angular momentum at a given unperturbed energy \eqref{E0def_app}, turning off all modes $a_{nm}$ with $n\neq m$. Although mode truncations are not usually possible in quantum mechanics due to quantum uncertainties, in this case, it follows immediately that the matrix elements \eqref{blockdef} between LLL and non-LLL states vanish. Indeed, $\mathcal{H}_{int}$ conserves the total angular momentum $M$ and particle number $N$ of a state, while first order perturbation theory perturbation requires both states in the overlap \eqref{blockdef} to have the same unperturbed energy $E$. Hence, the states involved in the diagonalization of a given sector are at fixed value of ($N$,$E$,$M$). The sector with $E = M$, in particular, contains the LLL states and no other state, since $m \leq n$ for any mode $a^\dagger_{nm}$. Hence, the LLL sector decouples from the other states in first order perturbation theory and the relevant Fock states become 
\beq
|\eta_0,\eta_1,\dots\rangle=\prod_{k=0}^\infty\frac{(a^\dagger_k)^{\eta_k}}{\sqrt{\eta_k!}}|0,0,0,\dots\rangle,\quad a_n \equiv a_{nn}.
\label{eq: LLL fock states_app}
\eeq
In the LLL sector \eqref{eq: LLL fock states_app}, the interaction Hamiltonian \eqref{eq: updated interaction Hamiltonian} is equivalent to the LLL Hamiltonian 
\begin{align}
\mathcal{H}_{LLL}=&\hspace{1mm}{\textstyle\frac12}\hspace{-2mm}\sum_{n+m = k+l } \hspace{-2mm}C_{nmkl}  {a}^{\dagger}_{n}{a}^{\dagger}_{m} {a}_{k}{a}_{l},
\label{eq: LLL Hamiltonian_app}
\end{align}
with
\begin{align}
C_{nmkl} = \frac{(n+m)!}{2^{n+m}\sqrt{n!m!k!l!}}.
\label{CLLL_app}
\end{align}

The conservation laws for the LLL Hamiltonian and their nonvanishing commutators are
\begin{align}
&\hspace{-1mm}{N} = \sum {a}^\dagger_k {a}_k, \,\,{M} = \sum k\,{a}^\dagger_k {a}_k,\,\,
{Z} = \sum\sqrt{k+1} \,a^\dagger_{k+1} a_k,\nn\\
&\hspace{-1mm}[{M}, {Z}] = {Z}, \quad [{M}, {Z}^\dagger]=-{Z}^\dagger, \quad
 [Z,Z^\dagger]=-N.\label{NEZcomm_app}
\end{align}
The operator $M$ corresponds to the angular momentum restricted to the LLL sector, while $N$ is the particle number operator which is manifestly conserved by the LLL Hamiltonian \eqref{eq: LLL Hamiltonian_app}. The conservation of the operator $Z$ is not obvious, but can be verified for the interaction coefficients \eqref{CLLL_app}, and comes from the decoupled center-of-mass motion in the harmonic potential.

With regard to the diagonalization problem for the Hamiltonian \eqref{eq: LLL Hamiltonian_app}, the Fock space separates into finite-dimensional, decoupled Hilbert spaces at fixed value of $N$ and $M$. The dimension of one block is determined by the number of ways $M$ units can be partitioned into at most $N$ parts. This number is denoted by $p_N(M)$ in number theory. 

Finally, we mention an additional structure controlling the energy eigenstates. It follows from the commutation relations \eqref{NEZcomm_app} that the $Z$-operator acts as a raising operator on energy eigenstates, moving states from block $(N,M)$ to block $(N,M+1)$ and copying the $p_N(M)$ eigenvalues of $\mathcal{H}_{LLL}$ from the lower to the higher block. The LLL Hamiltonian can be diagonalized simultaneously with $ZZ^\dagger$, and the eigenvalues of this latter operator keep track of the number of times the raising operator $Z$ has been applied to the eigenstate. In an $(N,M)$-block, the eigenvalues of $ZZ^\dagger$ can straightforwardly be deduced from the commutation relations \eqref{NEZcomm_app} and take values $l N$, with $l\in \{ 0,1,\cdots,M\}$. The eigenvalues that are not coming from lower blocks are associated with null vectors of $ZZ^\dagger$.

\section{Symmetries of the Husimi distribution}\label{appsymm}

In the main text, we stated that the Husimi distribution of an eigenstate of an operator bilinear in the creation-annihilation operators is invariant under the symmetry transformations generated by the corresponding classical conserved quantity. We demonstrate the validity of this claim for infinitesimal transformations. Start by considering the quadratic operator 
\beq
\label{eq: quadratic operator}
\hat{h} = \sum_n h(n) a^\dagger_n a_n,
\eeq
for an arbitrary function $h(n)$, and an eigenstate $|\psi\rangle$ of $\hat{h}$, so that $\hat{h}|\psi\rangle=h|\psi\rangle$. We aim to prove that
\beq
Q_{\psi}(\alpha) = Q_{\psi}(\alpha+\delta_{h} \alpha),
\label{eq: claim}
\eeq
where $\alpha+\delta_{h} \alpha$ is obtained by acting on the phase-space point $\alpha$ infinitesimally with the corresponding classical symmetry transformations $\alpha_n\to e^{i\lambd h(n)}\alpha_n$, so that 
\beq
\delta_{h}\alpha_n = i\lambd h(n)\alpha_n.
\eeq
To show \eqref{eq: claim}, we first note that, to linear order in $\lambd$, $\hat{h}$ can be inserted at no cost in the Husimi distribution
\beq
|\langle \psi|1+i\lambd\hat{h} |  \alpha \rangle|^2  = |\langle \psi|1+ i\lambd h |  \alpha \rangle|^2 =Q_{\psi}(\alpha),
\eeq
up to an irrelevant contribution at order $\lambd^2$, since $h$ is real. 
Next, we consider the action of $\hat{h}$ on $|  \alpha \rangle$ and make use of the Baker-Campbell-Hausdorff (BCH) identity
\beq
e^XYe^{-X} = Y + [X,Y]+\frac{1}{2!}[X,[X,Y]]+\cdots \label{eq: taylor}
\eeq
to push $\hat{h}$ through the exponential, with 
\beq
X = - \left( \sum_{i} \alpha_i a_i^\dagger-\alpha^*_i a_i \right) \hspace{0.35cm} \text{and}  \hspace{0.5cm} Y = 1+ i\lambd\hat{h}.
\eeq
Since $\hat{h}$ is quadratic in the modes, the Taylor series in \eqref{eq: taylor} terminates after the quadratic term. The remainder can be combined as
\begin{align*}
 &Q_{\psi}(\alpha) = |\langle \psi|1+ i\lambd\hat{h} |  \alpha \rangle|^2  \\
&= |\langle \psi |e^{-X}\left( 1+\sum_n ( a_n^\dagger\delta_{h}\alpha_n -a_n \delta_{h}\alpha_n^*-\alpha_n \delta_{h}\alpha^*_n)  \right)|0 \rangle|^2 \\
&= |\langle \psi |e^{-X}e^{\sum_n ( a_n^\dagger\delta_{h}\alpha_n -a_n \delta_{h}\alpha_n^*)}e^{-\sum_n \alpha_n \delta_{h}\alpha^*_n)}|0 \rangle|^2 \\
&= |\langle \psi |  e^{\sum_{n} (\alpha_n +\delta_{h}\alpha_n)a_n^\dagger-(\alpha^*_n +\delta_{h}\alpha^*_n) a_n}| 0 \rangle|^2 \\
&=|\langle \psi|  \alpha + \delta_{h}\alpha \rangle|^2 = Q_{\psi}(\alpha+\delta_{h} \alpha),
\end{align*}
where we used that $\hat{h}$ annihilates the vacuum in going from line 1 to 2, and used the BCH formula a second time to group the exponentials in going from line 3 to 4. This proves \eqref{eq: claim}. In the main text, our Hamiltonian commutes with both $N$ and $M$ so that the Husimi distribution of an energy eigenstate takes a constant value over each two-dimensional torus defined by these bilinears.

\section{Details of the sampling procedure}\label{appsampl}

For our visualization strategy, the first step consists in finding a phase-space point where the Husimi function is high. This will be the starting point of the Metropolis-Hastings sampling. Our method is slightly different for ladder and non-ladder states, since one can take advantage of the analytic understanding of the former states to speed up this part of the numerics.
For the ladder state, we expect the distribution to peak on the classical configurations given by  the ansatz
\begin{align}
\alpha_n= \left(\frac{a}{p}\,  n+b \right) \frac{p^n}{\sqrt{n!}}.
    \label{eq: ansatz_app}
\end{align}
The complex parameters $a$, $b$ and $p$ should be chosen so that the conserved quantities match the quantum numbers of the ladder state, see \cite{LLLres} for such relations. At any finite $N$, one notices that the maximum of the Husimi function is not quite located on the classical ansatz, which is understandable due to quantum uncertainties. However, further applying a local optimization algorithm on the ansatz point changes the Husimi function very little. Therefore, for ladder states, we choose the ansatz point as the initial point $\alpha_{\max}$ of the Metropolis-Hastings algorithm.
 For generic states, the starting point of the Metropolis-Hastings algorithm is generated using MATLAB's Global Optimization package applied on the Husimi function. Although finding the global maximum of a function numerically is a hard problem and the generated solution is not guaranteed to be correct, we found this algorithm to be useful to produce a phase-space point where the Husimi function is high (i.~e.~of a similar order of magnitude as for the ladder state).

Once a maximum of the Husimi distribution is found, we run the Metropolis-Hastings algorithm with $\alpha_{\max}$ as the seed and a Gaussian target distribution constraining the classical values $H$, $N$, $M$ and $|Z|^2$ of the sampled points around their values at $\alpha_{\max}$. 
Our choice of the target distribution allows the Metropolis-Hastings algorithm to wander in phase space while remaining close to the shell of constant conservation laws $M_\psi$.
The classical energy of the phase-space point where the Husimi distribution peaks is generally expected to be within quantum 
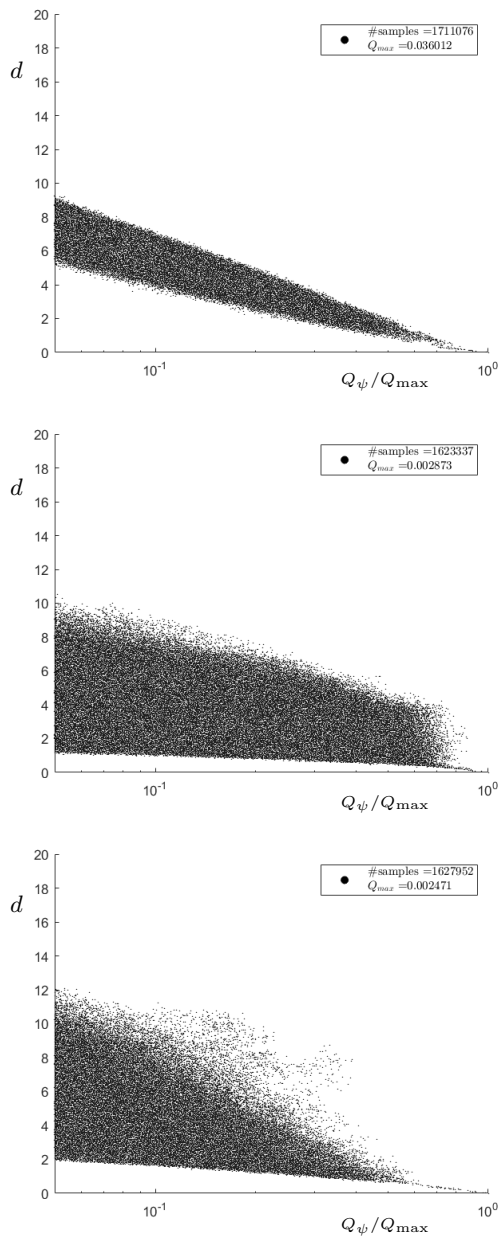
\begin{figure}[H]\vspace{-5mm}
\centering
    \includegraphics[scale=0.4]{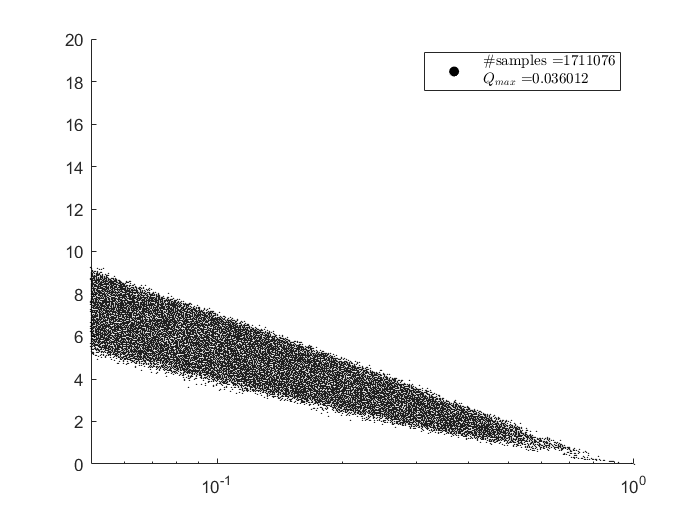}\\
    \includegraphics[scale=0.4]{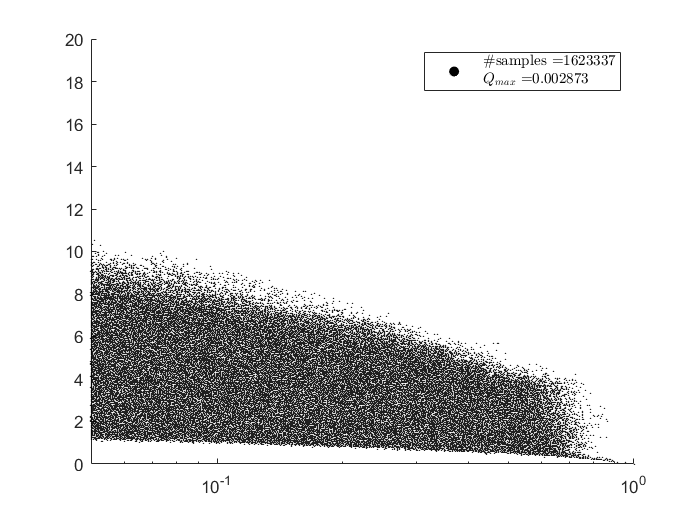}\\
    \includegraphics[scale=0.4]{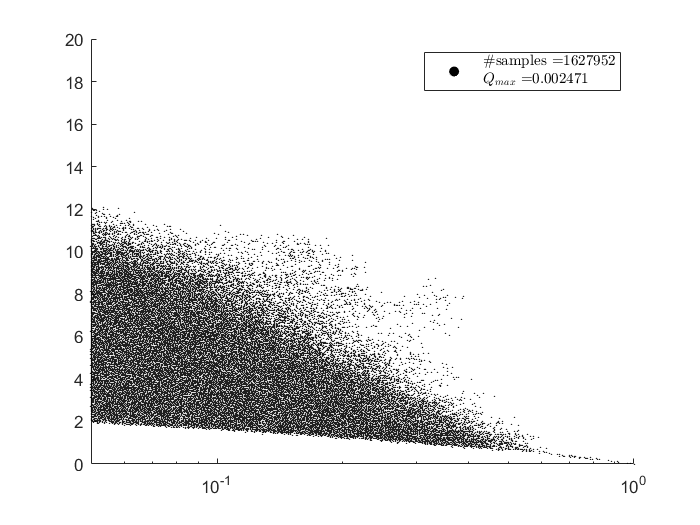}\\
   \begin{picture}(0,0)
     \put(30,335){$\scriptstyle Q_\psi/Q_{\max}$}
     \put(30,175){$\scriptstyle Q_\psi/Q_{\max}$}
     \put(30,15){$\scriptstyle Q_\psi/Q_{\max}$}
     \put(-95,135){$d$}
     \put(-95,293){$d$}
     \put(-95,451){$d$}
   \end{picture}
    \caption{Husimi function values over a cloud of phase-space points against the distance $d$ to the initial two-dimensional surface $\Theta(\alpha_{\max})$. \textbf{Top:} for the ladder state with $|Z|=0$, $N=30$ and $M=30$. \textbf{Middle:} for a non-ladder state with $|Z|=0$, $N=30$ and $M=30$. \textbf{Bottom:} for the non-ladder ground state of the Hilbert space block with $N=20$ and $M=30$. The Gaussian distribution used in the sampling algorithm has mean values determined by the classical values of the initial optimized phase-space point $(N_{\max},M_{\max},Z_{\max},H_{\max})=(30,30,0,225)$ (top), $(N_{\max},M_{\max},Z_{\max},H_{\max})=(30,30,0,240.32)$ (middle) and $(N_{\max},M_{\max},Z_{\max},H_{\max})=(20,30,0,84.60)$ (bottom). The standard deviations of the Gaussian distribution are set to $5\sqrt{2} \times(\sigma_N,\sigma_M,\sigma_{|Z|^2},\sigma_H) = (1,1,N,N)$ for all three samplings. The three plots show again that the Husimi distribution is only well-localized on $\Theta(\alpha_{\max})$ when the considered eigenstate is a ladder state. Such localization is not observed when the ground state is not a ladder state. }
    \label{fig:nonladder}
\end{figure}
\noindent 
uncertainty of the quantum eigenstate energy. (Some differences between classical and quantum quantities may be attributed to ordering ambiguities.) The classical values of the other three conserved quantities, besides the Hamiltonian, turn out to be exactly equal to the corresponding quantum number.
Regarding the standard deviations of the Gaussian distribution, one needs to compromise between a deviation large enough allowing the algorithm to efficiently move away from the two-dimensional torus  $\Theta(\alpha_{\max})$ during the sampling, but also small enough to eliminate contamination of the samples with points where the Husimi distribution takes negligible values that would not provide any useful information on the topography of the Husimi distribution. 
Since the Hamiltonian and $ZZ^\dagger$ operator are quartic in the modes, we let the standard deviations $\sigma_H$ and $\sigma_{|Z|^2}$ of the Gaussian distribution scale with $N$, while $\sigma_{N}$ and $\sigma_{M}$ are fixed at order 1, corresponding to the magnitude of typical ordering ambiguities in the corresponding operators.
For definiteness, we chose the values $5\sqrt{2} \times(\sigma_N,\sigma_M,\sigma_{|Z|^2},\sigma_H) = (1,1,N,N)$.
In the Metropolis-Hastings sampling, the $(n\!+\!1)^\text{th}$ candidate sample point is generated by a (multivariate) Gaussian distribution, with the $n^\text{th}$ sample point taken as its mean. The standard deviation of this Gaussian is a measure for the step size to the next candidate, and was chosen to be 
$10^{-3}$.\ \  For numerical efficiency,\ \  we only vary the lowest
10 classical modes when moving from one sample to the next. This truncation does not significantly affect our results, as long as the modes $a_{n\geq10}$ have small expectation values in the quantum state of interest. 

\section{No localization for non-ladder ground states}\label{appnonladd}

At $M>N$, ground states of the LLL system are not necessarily ladder states anymore. It is therefore interesting to compare the Husimi distributions of ladder and non-ladder ground states. As an illustration, we perform the same analysis as done in the main text on 
\begin{enumerate}
\item[(1)] the ground state in the ($30,30$)-block, which is a ladder state, 
\item[(2)] a generic state in the ($30,30$)-block, 
\item[(3)] the ground state in the ($20,30$)-block, which is not a ladder state. 
\end{enumerate}
(All of these states are annihilated by $Z^\dagger$.)
The dimension of the $(30,30)$-block is $p_N(M) = 5604$, while the $(20,30)$-block has dimension $p_N(M) = 5507$. For the ($30,30$)-block, the ground state is still a ladder state and the top two plots of Fig.~\ref{fig:nonladder} show that the same differences between the ground state and a generic state can be observed as in the main text. For the ground state, the only phase-space points with large Husimi distribution are located in the vicinity of the two-dimensional surface $\Theta(\alpha_{\max})$. By contrast, for the generic eigenstate, we observe large Husimi values at phase-space points far away from $\Theta(\alpha_{\max})$.

For the ($20,30$)-block, the ground state is no longer a ladder state. In this case, no localization of the Husimi distribution is observed, as seen on the bottom plot of Fig.~\ref{fig:nonladder}, where a swarm of points appears on the right of the main cloud, showing that there are points in phase space far away from the initial surface $\Theta(\alpha_{\max})$ where the Husimi function is nevertheless large. Thus, the non-ladder state is less tightly bound and less well-localized, compared to ladder states, despite being at the bottom of the spectrum.

\onecolumngrid


\vspace{10mm}
\begin{thebibliography}{99}\vspace{5mm}
\twocolumngrid

\bibitem{Eckhardt:1995}
B.~Eckhardt, S.~Fishman, J.~Keating, O.~Agam, J.~Main and K.~M\"uller,
{\it Approach to ergodicity in quantum wave functions,}
\doi{Phys. Rev. E \textbf{52} (1995) 5893}{10.1103/PhysRevE.52.5893},
\arXiv{chao-dyn/9509017}.

\bibitem{Serbyn:2021}
M.~Serbyn, D.~Abanin and Z.~Papi\'c,
{\it Quantum many-body scars and weak breaking of ergodicity,}
\doi{Nature Phys. \textbf{17} (2021) 675-685}{10.1038/s41567-021-01230-2},
\arXiv{2011.09486} [quant-ph].

\bibitem{Vikram:2022}
A.~Vikram and V.~Galitski,
{\it Dynamical quantum ergodicity from energy level statistics,}
\doi{Phys. Rev. Res. \textbf{5} (2023) 033126}{10.1103/PhysRevResearch.5.033126},
\arXiv{2205.05704} [quant-ph].

\bibitem{Bloch:2008zzb}
I.~Bloch, J.~Dalibard and W.~Zwerger,
{\it Many-body physics with ultracold gases,}
\doi{Rev. Mod. Phys. \textbf{80} (2008) 885}{10.1103/RevModPhys.80.885},
\arXiv{0704.3011} [cond-mat.other].

\bibitem{Fetter:2009zz}
A.~L.~Fetter,
{\it Rotating trapped Bose-Einstein condensates,}
\doi{Rev. Mod. Phys. \textbf{81} (2009) 647}{10.1103/RevModPhys.81.647},
\arXiv{0801.2952} [cond-mat.stat-mech].

\bibitem{Ho:2001zz}
T.-L.~Ho,
{\it Bose-Einstein condensates with large number of vortices,}
\doi{Phys. Rev. Lett. \textbf{87} (2001) 060403}{10.1103/PhysRevLett.87.060403},
\arXiv{cond-mat/0104522}.

\bibitem{ABD}A.\ Aftalion, X.\ Blanc and J.\ B.\ Dalibard, {\it Vortex patterns in a fast rotating Bose-Einstein condensate,}
\doi{Phys.\ Rev.\ A {\bf 71} (2005) 023611}{10.1103/PhysRevA.71.023611}, \arXiv{cond-mat/0410665}.

\bibitem{ABN}A.~Aftalion, X.~Blanc and F.~Nier, {\it Lowest Landau level functional and Bargmann spaces
for Bose-Einstein condensates}, \doi{J.\ Func.\ Anal. 241 (2006) 661}{10.1016/j.jfa.2006.04.027}.

\bibitem{Husimi}
K.~Husimi,
{\it Some formal properties of the density matrix,}
\doi{Proc. Phys. Math. Soc. Jpn. \textbf{22} (1940) 264}{10.11429/ppmsj1919.22.4_264}.

\bibitem{rev}H.-W.~Lee,
{\it Theory and application of the quantum phase-space distribution functions},
\doi{Phys.\ Rep. {\bf 259} (1995) 147}{10.1016/0370-1573(95)00007-4}.

\bibitem{phasespaceloc1}P.~Leboeuf, J.~Kurchan, M.~Feingold and D.~P.~Arovas,
{\it Phase-space localization: topological aspects of quantum chaos,}
\doi{Phys.\ Rev.\ Lett. {\bf 65} (1990) 3076}{10.1103/PhysRevLett.65.3076}.

\bibitem{phasespaceloc2}
S.~Nonnenmacher and A. Voros,
{\it Chaotic eigenfunctions in phase space}, \doi{J.\ Stat.\ Phys. {\bf 92} (1998) 431}{10.1023/A:1023080303171}.

\bibitem{phasespaceloc3}
C.~Aulbach, A.~Wobst, G.-L.~Ingold, P.~H\"anggi and I.~Varga, {\it Phase-space visualization of a metal–insulator transition},
\doi{New\ J.\ Phys. {\bf 6} (2004) 70}{10.1088/1367-2630/6/1/070}.

\bibitem{phasespaceloc4}B.~Batisti\'c and M.~Robnik,
{\it Quantum localization of chaotic eigenstates and the level spacing distribution},
\doi{Phys.\ Rev.\ E {\bf 88} (2013) 052913}{10.1103/PhysRevE.88.052913}, \arXiv{1310.2483} [quant-ph].

\bibitem{phasespaceloc5}
S.~Pilatowsky-Cameo, D.~Villase\~nor, M.~A.~Bastarrachea-Magnani, S.~Lerma-Hern\'andez, L.~F.~Santos, J.~G.~Hirsch,
{\it Ubiquitous quantum scarring does not prevent ergodicity,}
\doi{Nature Comm. \textbf{12} (2021) 852}{10.1038/s41467-021-21123-5}, \arXiv{2009.00626} [cond-mat.stat-mech];
{\it Identification of quantum scars via phase-space localization measures,}
\doi{Quantum \textbf{6} (2022) 644}{10.22331/q-2022-02-08-644}, \arXiv{2107.06894} [quant-ph].

\bibitem{phasespaceloc6}Q.~Hummel, K.~Richter and P.~Schlagheck,
{\it Genuine many-body quantum scars along unstable modes in Bose-Hubbard systems},
\doi{Phys.\ Rev.\ Lett. {\bf 130} (2023) 250402}{10.1103/PhysRevLett.130.250402},\\ \arXiv{2212.12046} [cond-mat.quant-gas].

\bibitem{phasespaceloc7}
Q.~Wang and M.~Robnik,
{\it Statistics of phase space localization measures and quantum chaos in the kicked top model,}
\doi{Phys. Rev. E \textbf{107} (2023) 054213}{10.1103/PhysRevE.107.054213}
\arXiv{2303.05216} [nlin.CD].

\bibitem{split}
B.~Craps, M.~De Clerck, O.~Evnin and S.~Khetrapal,
{\it Energy level splitting for weakly interacting bosons in a harmonic trap,}
\doi{Phys. Rev. A \textbf{100} (2019) 023605}{10.1103/PhysRevA.100.023605},\\
\arXiv{1903.04974} [cond-mat.quant-gas].

\bibitem{split1}T.~Busch, B.-G. Englert, K.~Rz\k{a}\.zewski and M.~Wilkens, \emph{Two cold atoms in a harmonic trap,} \doi{Found. Phys. {\bf 28} (1998) 549}{10.1023/A:1018705520999}.

\bibitem{split2}M.~Block and M.~Holthaus, \emph{Pseudopotential approximation in a harmonic trap,} \doi{Phys.\ Rev.\ A {\bf 65} (2002) 052102}{10.1103/PhysRevA.65.052102}.

\bibitem{split3}F.~Werner and Y.~Castin,  \emph{The unitary three-body problem in a trap,} \doi{Phys.\ Rev.\ Lett.\ {\bf 97} (2006) 150401}{10.1103/PhysRevLett.97.150401}, \arXiv{cond-mat/0507399}; \emph{The unitary gas in an isotropic harmonic trap: symmetry properties and applications,} \doi{Phys.\ Rev.\ A {\bf 74} (2006) 053604}{10.1103/PhysRevA.74.053604}, \arXiv{cond-mat/0607821}.

\bibitem{split4}P.~Shea, B.~P.~van Zyl and R.~K.~Bhaduri,  \emph{The two-body problem of ultra-cold atoms in a harmonic trap,} \doi{Am. J. Phys. {\bf 77}  (2009) 511}{10.1119/1.3013812}, \arXiv{0807.2979} [physics.atom-ph].

\bibitem{split5}N.~L.~Harshman, \emph{Symmetries of three harmonically trapped particles in one dimension}, \doi{Phys. Rev. A {\bf 86} (2012) 052122}{10.1103/PhysRevA.86.052122}, \arXiv{1209.1398} [quant-ph].

\bibitem{split6}M.~A.~Garc\'\i a-March, B.~Juli\'a-D\'\i az, G.~E.~Astrakharchik, J.~Boronat and A. Polls, \emph{Distinguishability, degeneracy and correlations in three harmonically trapped bosons in one-dimension,} \doi{Phys.\ Rev.\ A {\bf 90} (2014) 063605}{10.1103/PhysRevA.90.063605},\\ \arXiv{1410.7307} [cond-mat.quant-gas].

\bibitem{split7}P.~Mujal, E.~Sarl\'e, A.~Polls and B.~Juli\'a-D\'\i az, \emph{Quantum correlations and degeneracy of identical bosons in a two-dimensional harmonic trap,} \doi{Phys.\ Rev.\ A {\bf 96} (2017) 043614}{10.1103/PhysRevA.96.043614}, \arXiv{1707.04166} [cond-mat.quant-gas].

\bibitem{split8}P.~Ko\'scik and T.~Sowi\'nski, \emph{Exactly solvable model of two trapped quantum particles interacting via finite-range soft-core interactions,} \doi{Sci.\ Rep. {\bf 8} (2018) 48}{10.1038/s41598-017-18505-5}, \arXiv{1707.04240} [cond-mat.quant-gas].

\bibitem{split9}D.~Blume, M.~W.~C.~Sze and  J.~ L.~Bohn, \emph{Harmonically trapped four-boson system,} \doi{Phys.\ Rev.\ A {\bf 97} (2018) 033621}{10.1103/PhysRevA.97.033621}, \arXiv{1802.00129} [cond-mat.quant-gas].

\bibitem{split10}
N.~Z.~Lizama, S.~C.~Carrasco, J.~Rogan and J.~A.~Valdivia,
{\it Three-dimensional non-approximate Coulomb interaction between two trapped quantum particles,}
\doi{Sci. Rep. \textbf{13} (2023) 18210}{10.1038/s41598-023-45234-9}.

\bibitem{GHT}P.~Germain, Z.~Hani and L.~Thomann, {\it On the continuous resonant equation for NLS: I. Deterministic analysis,}
\doi{J.\ Math.\ Pur.\ App.\ {\bf 105} (2016) 131}{10.1016/j.matpur.2015.10.002}, \arXiv{1501.03760} [math.AP].

\bibitem{GT}P.~Germain and L.~Thomann, {\it On the high frequency limit of the LLL equation,} \doi{Quart.\ Appl.\ Math. {\bf 74} (2016) 633}{10.1090/qam/1435}, \arXiv{1509.09080} [math.AP].

\bibitem{LLLres}A.~Biasi, P.~Bizo\'n, B.~Craps and O.~Evnin,
{\it Exact lowest-Landau-level solutions for vortex precession in Bose-Einstein condensates,}
\doi{Phys. Rev. A \textbf{96} (2017) 053615}{10.1103/PhysRevA.96.053615},\\
\arXiv{1705.00867} [cond-mat.quant-gas].

\bibitem{LLL2}
P.~G\'erard, P.~Germain and L.~Thomann,
{\it On the cubic lowest Landau level equation,}
\doi{Arch. Rat. Mech. Anal. \textbf{231} (2019) 1073}{10.1007/s00205-018-1295-4},
\arXiv{1709.04276} [math.AP].

\bibitem{LLL3}
P.~Germain, V.~Schwinte and L.~Thomann,
{\it On the stability of the Abrikosov lattice in the lowest Landau level,}
\arXiv{2404.06085} [math.AP].
  
\bibitem{breathing}
  O.~Evnin,
  {\em Breathing modes, quartic nonlinearities and\\ effective resonant systems,}
\doi{SIGMA {\bf 16} (2020) 034}{10.3842/SIGMA.2020.034},\\ \arXiv{1912.07952} [math-ph].

\bibitem{quantres}
O.~Evnin and W.~Piensuk,
{\it Quantum resonant systems,\\ integrable and chaotic,}
\doi{J. Phys. A \textbf{52} (2019) 025102}{10.1088/1751-8121/aaf2a1},\\
\arXiv{1808.09173} [math-ph].

\bibitem{ladder}
M.~De Clerck and O.~Evnin,
{\it Time-periodic quantum states of weakly interacting bosons in a harmonic trap,}
\doi{Phys. Lett. A \textbf{384} (2020) 126930}{10.1016/j.physleta.2020.126930},
\arXiv{2003.03684} [cond-mat.quant-gas].

\bibitem{largec} 
B.~Craps, M.~De Clerck and O.~Evnin,
{\it Time-periodicities in holographic CFTs,}
\doi{JHEP \textbf{09} (2021) 030}{10.1007/JHEP09(2021)030},
\arXiv{2103.12798} [hep-th].

\bibitem{Schwinte}
V.~Schwinte,
{\it An optimal minimization problem in\\ the lowest Landau level and related questions,}
\doi{Comm. Math. Phys. \textbf{405} (2024) 98}{10.1007/s00220-024-04974-z},
\arXiv{2303.03902} [math.AP].

\bibitem{HusWig}
M.~Hillery, R.~F.~O'Connell, M.~O.~Scully and E.~P.~Wigner, {\it Distribution functions in physics: fundamentals}, 
\doi{Phys.\ Rep. {\bf 106} (1984) 121}{10.1016/0370-1573(84)90160-1}.

\bibitem{decay2}
K.~Takahashi, {\it Wigner and Husimi functions in quantum mechanics,}
\doi{J.\ Phys.\ Soc.\ Jpn. \textbf{55} (1986) 762}{10.1143/JPSJ.55.762}.

\bibitem{decay1}
J.~Kurchan, P.~Leboeuf and M.~Saraceno, {\it Semiclassical approximations in the coherent-state representation,}
\doi{Phys. Rev. A \textbf{40} (1989) 680}{10.1103/PhysRevA.40.6800}.

\bibitem{codes} All MATLAB codes used for our numerical work are available from \doi{DOI:10.5281/zenodo.12704876}{10.5281/zenodo.12704876}.

\bibitem{extralddr}It was pointed out to us by Sara Vanovac that patterns similar to ladder states exist in some nonintegrable spin chains, see\\
S.~Moudgalya, S.~Rachel, B.~A.~Bernevig and N.~Regnault, {\it Exact excited states of nonintegrable models,} 
\doi{Phys.\ Rev.\ B {\bf 98} (2018) 235155}{10.1103/PhysRevB.98.235155}, \arXiv{1708.05021} [cond-mat.str-el];\\
D.~K.~Mark, Ch.-J. Lin and O.~I.~Motrunich, {\it Unified structure for exact towers of scar states in the AKLT and other models},
\doi{Phys. Rev. B {\bf 101} (2020) 195131}{10.1103/PhysRevB.101.195131}, \arXiv{2001.03839} [cond-mat.str-el].

\bibitem{firstq}While this paper was in final stages of preparation, Nicolas Rougerie drew our attention to the construction of ladder states in terms of the first-quantized representation in\\
G.~F.~Bertsch and T.~Papenbrock, {\it Yrast line for weakly interacting trapped bosons,}
\doi{Phys.\ Rev.\ Lett. {\bf 83} (1999) 5412}{10.1103/PhysRevLett.83.5412}, \arXiv{cond-mat/9908189};\\
R.~A.~Smith and N.~K.~Wilkin, {\it Exact eigenstates for repulsive bosons in two dimensions,}
\doi{Phys.\ Rev.\ A {\bf 62} (2000) 061602(R)}{10.1103/PhysRevA.62.061602}, \arXiv{cond-mat/0005230}.\\
The first-quantized wavefunctions are expressed through simple symmetric polynomials of the particle coordinates, which does not suggest any form of localization in the corresponding particle phase space.

\bibitem{Qopt}
W.~P.~Schleich, {\it Quantum optics in phase space} (Wiley-VCH, 2001).

\bibitem{part_gen} 
D.~Holdaway, {\it Integer partition generator,} MATLAB Central File Exchange, retrieved
February 28, 2022. \url{https://www.mathworks.com/matlabcentral/fileexchange/36437-
integer-partition-generator}.

\end{thebibliography}
\end{document}